\documentclass[prl]{revtex4}
\usepackage{graphicx}
\usepackage{bm}
\begin{document}
\newcommand{\alphatil}{\frac{1}{\tau}}
\newcommand{\jtil}{\tilde{j}}
\newcommand{\Kp}{{K_\perp}}
\newcommand{\Mw}{{M_w}}
\newcommand{\omegaz}{{\omega_0}}
\newcommand{\Omegaz}{{\Omega_0}}
\newcommand{\Rtil}{\tilde{R}}
\newcommand{\tauz}{{\tau_0}}
\newcommand{\Tv}{{\bm T}}
\newcommand{\nv}{{\bm n}}
\newcommand{\Ne}{{N_{\rm e}}}
\newcommand{\eF}{{\epsilon_F}}
\newcommand{\Ev}{{\bm E}}
\newcommand{\kF}{{k_F}}
\newcommand{\kp}{{k_\perp}}
\newcommand{\sigmav}{{\bm \sigma}}
\newcommand{\phiz}{{\phi_0}}
\newcommand{\rhow}{{\rho_{\rm w}}}
\newcommand{\Vz}{{V_0}}
\newcommand{\Mv}{{\bm M}}
\newcommand{\tiln}{{\tilde{n}}}
\newcommand{\tilv}{{\tilde{v}}}
\newcommand{\tilK}{{\tilde{K}_\perp}}
\newcommand{\tz}{{t_0}}
\newcommand{\Area}{A}
\newcommand{\dx}{{d^3 x}}
\newcommand{\xv}{{\bm x}}
\newcommand{\rv}{{\bm r}}
\newcommand{\kv}{{\bm k}}
\newcommand{\qv}{{\bm q}}
\newcommand{\Vv}{{\bm V}}
\newcommand{\vv}{{\bm v}}
\newcommand{\Av}{{\bm A}}
\newcommand{\Bv}{{\bm B}}
\newcommand{\av}{{\bm a}}
\newcommand{\DOS}{{\nu}}
\newcommand{\kB}{{k_B}}
\newcommand{\js}{{j_{\rm s}}}
\newcommand{\jsc}{{j_{\rm s}^{\rm cr}}}
\newcommand{\jsca}{{j_{\rm s}^{\rm cr (1)}}}
\newcommand{\jscb}{{j_{\rm s}^{\rm cr (2)}}}
\newcommand{\jc}{{j^{\rm cr}}}

\title{ 
Domain wall displacement triggered by an 
AC current below threshold
}
\author{Gen Tatara}
\affiliation{
PRESTO, JST, 4-1-8 Honcho Kawaguchi, Saitama, Japan \\
Graduate School of Science, Osaka University, Toyonaka, Osaka 560-0043, 
Japan}

\author{Eiji Saitoh} 
\affiliation{
Department of Physics, Keio University, Yokohama 223-8522, Japan
}

\author{Masahiko Ichimura} 
\affiliation{
Advanced Research Laboratory, Hitachi, Ltd., 
Hatoyama, Saitama, 350-0395, Japan
}

\author{Hiroshi Kohno} 
\affiliation{Graduate School of Engineering Science, 
Osaka University, Toyonaka, Osaka 560-8531, Japan
}

\date{\today}

\begin{abstract}
It is theoretically demonstrated that a displacement of a
pinned domain wall, typically of order of $\mu$m, can be 
driven by use of an ac current which is below threshold value.
The point here is that finite motion around the pinning center by a low
 current is enhanced significantly by the resonance 
if the frequency is tuned close to the pinning frequency as demonstrated by recent experiment.
\end{abstract}
\maketitle

Domain wall motion driven by electric current is of special interest
recently. From the viewpoint of application to magnetic memories, most
urgent issue is to reduce the threshold current. 
Threshold current in experiments so far on metallic wires under DC or
slow pulse
current is mostly of order of $10^{11}\sim 10^{12}$[A/m$^2$]
\cite{Grollier02,Vernier03,Klaeui03,Yamaguchi04}.
Smaller value of $10^{10}$[A/m$^2$] is reported when a pulse of order of
ns is applied\cite{Lim04}.

To realize lower current density, further understanding of the driving
mechanism as well as extrinsic effects such as pinning are necessary.
Current-driven domain wall motion in an artificial pinning potential
was studied quite recently\cite{Saitoh04}.
Besides the pinning potential being controllable, the experiment has a
novelty in that the applied current is ac current of MHz range, and the
resonance behavior was observed.
The data indicated surprisingly that the domain wall was driven mostly
by the force due to charge current, in contrast to the case of dc
current, in which the spin torque due to spin current dominates.
This was argued there to be due to the strong enhancement of the force at the
resonance and, at the same time, the reduction of spin torque effect 
in slow ac dynamics (compared with microscopic spin frequency ($\sim$GHz)).
The displacement  observed there is
driven below the critical current and thus is finite, but still it is
rather big (estimated to be $\sim 10\mu$m).
This result indicates that by use of ac field, domain wall can be
driven by a different mechanism from the dc case, 
 and this would be useful
in realizing domain wall displacement at low current.

Motivated by this experiment, we study in this paper the depinning of domain wall under ac
current theoretically. It is shown that depinning under ac field occurs
at current density which is lower
than the dc case roughly by a factor of Gilbert damping.
Thus use of ac current is quite promising for device application.

We consider a wire in $z$ direction. Choosing a hard axis as
$y$-direction, the spin Hamiltonian is
\begin{equation}
H_S= \frac{S^2}{2}
    \Big\{ J \big( (\nabla\theta)^2 + \sin^2 \theta(\nabla\phi)^2 \big)
   + \sin^2\theta \, ( K + K_\perp \sin^2\phi ) \Big\} \bigg], 
   \label{HS}
\end{equation}
where the easy- and hard-axis anisotropies ($K$ and $\Kp$) include the
effect of demagnetizing field. 
We consider a case of large $\Kp$ or small current (smaller than the
critical current), 
thus $\phi$ can be treated as small.
Note that most DC experiments so far is in the current region larger
than the critical value, thus the analysis below does not apply.
We consider a pinning potential of a harmonic type with a range of $\xi$,
\begin{equation}
V(X)=\frac{1}{2} \Mw\Omegaz^2X^2\theta(\xi-|X|)
\end{equation}
where $\Mw\equiv\hbar^2N/\Kp\lambda^2$ is the mass of wall, $\lambda\equiv\sqrt{J/K}$ is the wall thickness,  
and the oscillation
frequency at the bottom is denoted by $\Omegaz$.
We consider a current with frequency $\omegaz$.
The equation of motion for the domain wall at $z=X(t)$ is 
written as (we neglect terms of $O(\alpha^2)$,
where $\alpha$ is the Gilbert damping parameter.
)
\begin{equation}
\ddot{X}
+\alphatil \dot{X}
+\Omegaz^2 X 
=
f_j(t)
\end{equation}
where 
$\alphatil\equiv\alpha\frac{\Kp}{\hbar}
 \left(1+\left(\frac{\hbar\Omegaz}{\Kp}\right)^2\right)$.
$f_j$ is  the total force due to ac current  divided by $\Mw$, which
is given in terms of charge current $j$ and spin current $j_s$ 
as\cite{TK04}
\begin{equation}
f_j(t)
\equiv
\frac{a^3}{2Se}\left[ \frac{\Kp\lambda An}{\hbar}\frac{e^2 R_w}{\hbar}
\left(j+\alpha\frac{\hbar}{\Kp}\frac{\partial}{\partial t}{j}\right)
+\gamma \frac{\partial}{\partial t}{j}_s\right],
\label{eqX}
\end{equation}
where $\gamma$ is the adiabaticity parameter ($\gamma\rightarrow1$ for
thick wall, $\lambda k_F \gg 1$).
We consider a current with angular frequency of $\omegaz$ 
and the amplitude $j_0$ switched at $t=0$;
$j(t)=j_0 e^{-i\omegaz t}\theta(t)$. 
By using parameter $\beta$ representing the spin polarization of current
($j_s\equiv \beta j$), we can write 
 $f_j(t)= f_0(\omegaz)e^{-i\omegaz t}\theta(t)
+f_1 \delta(t) $, where
\begin{equation}
f_0(\omegaz)
\equiv
\frac{a^3}{2Se}\left[ \frac{\Kp}{\hbar}\Rtil
\left(1-i\alpha
\frac{\hbar\omegaz}{\Kp}\right)-i\beta\gamma \omegaz\right]j_0,
\label{f0}
\end{equation}
and $\Rtil\equiv \lambda A n\frac{e^2 R_w}{\hbar}$.
The $\delta$-function part is given by
\begin{equation}
f_1
\equiv
\frac{a^3}{2Se}\left( \alpha\Rtil+\beta\gamma \right)j_0.
\label{f1}
\end{equation}
We here neglect $f_1$, focusing on the resonating behavior due to
oscillating force, $f_0$.
The oscillating component of the solution is then obtained as
\begin{equation}
X(t)=\frac{f_0}{2\Omega}e^{-i\omegaz t}
\left(
\frac{1-e^{[-\frac{1}{2\tau}+i(\omegaz+\Omega)]t}}
{(\omegaz+{\Omega}+i\frac{1}{2\tau})}
-
\frac{1-e^{[-\frac{1}{2\tau}+i(\omegaz-\Omega)]t}}
{(\omegaz-{\Omega}+i\frac{1}{2\tau})}
\right)
\end{equation}
where
$\Omega\equiv 
\sqrt{\Omegaz^2-\frac{1}{4\tau^2}}$.
After a long time, $t\gtrsim \tau$,
the amplitude of $X$ reduces to  
\begin{equation}
|X(t)|\rightarrow |f_0| \frac{1}
{\sqrt{ (\omegaz^2-\Omegaz^2)+\alphatil^2\omegaz^2 }}
\end{equation}
This is enhanced at resonance, $\omegaz=\Omegaz$, to be
\begin{equation}
|X(t)|_{\rm res}= |f_0| \frac{1}
{\alphatil\Omegaz}
=\frac{\hbar |f_0|} 
{\alpha \Kp \Omegaz (1+(\frac{\Omegaz}{\Kp})^2)}
\end{equation}
Depinning occurs when $|X(t)|\gtrsim \xi$.
At resonance, the condition is given by
$|f_0|=\alphatil\Omegaz\xi
=\alpha \Kp \Omegaz(1+(\frac{\Omegaz}{\Kp})^2) /{\hbar } 
\equiv f_{\rm c}^{\rm ac}$.
The threshold current is 
\begin{equation}
j_{\rm c}^{\rm ac}
=\frac{2Se}{a^3}\alpha\xi\Omegaz
\frac{\left(1+\left(\frac{\hbar\Omegaz}{\Kp}\right)^2\right)}
{\sqrt{
\Rtil^2
+\left(\frac{\hbar\Omegaz}{\Kp}\right)^2 
(\gamma \beta+\alpha\Rtil)^2}}
\label{jac}
\end{equation}
In the case of dc current, threshold condition is given by
$f_0= \Omegaz^2\xi\equiv f_{\rm c}^{\rm dc}$, i.e., 
$f_{\rm c}^{\rm ac}=\frac{1}{\tau\Omegaz}f_{\rm c}^{\rm dc}$.
Thus the threshold current in the ac case at resonance is 
written in terms of dc case as 
$j_{\rm c}^{\rm ac}=\epsilon j_{\rm c}^{\rm dc}$, where
\begin{equation}
\epsilon \equiv \frac{1}{\tau \Omegaz}
=\alpha
\frac{\Kp}{\hbar \Omegaz}
 \left(1+\left(\frac{\hbar\Omegaz}{\Kp}\right)^2\right)
\end{equation}
Since $\alpha$ is generally small ($\sim 0.01$), we see that
depinning threshold is much lower when an ac current is applied than the
dc case unless $\Kp/\hbar\Omegaz$ is very large or very small. 
(In most metallic wires, the pinning angular frequency $\Omegaz$ is small 
compared with oscillation frequency scale of a single spin, $\Kp/\hbar$.)
We here note, however, that even in the case of very weak pinning created by
external magnetic field ($\Omegaz\sim 2\pi\times 25$MHz and $\Kp\sim 0.1$K) 
in Ref. \cite{Saitoh04}, the ac enhancement by a
factor of 2 is expected; $\epsilon \simeq 0.5$.
Displacement there is estimated to be 
$\Delta X \sim (a^3 j_0/2e)(\Rtil/\alpha \lambda \Omegaz) \sim 10\mu$m.

The dominant driving mechanism under ac current is determined by the ratio
(see eq. (\ref{jac}))
\begin{equation}
\eta=\beta\gamma \frac{\hbar \Omegaz}{\Rtil \Kp},
\end{equation}
where the spin transfer effect dominates over the momentum transfer 
if $\eta >1$.
In the case of Ref. \cite{Saitoh04}, $\Omegaz\simeq 2\pi\times 25$MHz, 
$\lambda=70$nm,
$A=70\times 45$nm$^2$. $\Kp$ is estimated to be 0.1K, and $a\sim 2.5$\AA.
Thus spin transfer dominates if 
$R_w\lesssim 0.5\beta\gamma\times 10^{-5} $[$\Omega$]. 
The spectrum obtained there suggests a larger resistance, 
$R_w\sim 3 \times 10^{-4} $[$\Omega$], i.e., 
momentum transfer is dominant, in sharp contrast to the
case of dc current. 
This experimental finding for a
thick wall with thickness $\sim70$nm is quite surprising.
At the same time, this fact has an important implication for applications. 
In fact, domain wall resistance decays quite rapidly for a thickness
larger than Fermi wavelength\cite{Cabrera74,TZMG99}, typically
exponentially $R_w\propto e^{-2\pi \zeta k_F\lambda}$ if the wall
profile is the form of $\tanh$, where $\zeta$ is
a number of order of unifty. In the case of linear wall, where $S_z$
changes linearly as function of $x$, it decays slower as
 $R_w\propto \frac{\sin^2 \zeta k_F \lambda}{ (k_F \lambda)^2}$.
Even in this modest case,we will have $R_w$
larger by an order of magnitude
if we can reduce the wall thickess by a factor of three, 
 and then the driving force would be larger by order of magnitude. 
For the $\tanh$-type wall, the change would be more drastic. 
It is critically important that for the ac case, where momentum transfer
contributes,  $R_w$ is another parameter to control the threshold.


For the application, switching speed is crucially important.
The speed based on resonance swithcing considered above is determined by 
the time scale $\tau$. 
In the case of Ref. \cite{Saitoh04}, 
($\Kp=0.1$K and $\alpha=0.01$), the time scale corresponds to
20MHz.
Faster operation would be possible by using  larger $\Kp$.

To implement the AC current -driven domain wall motion into device, one would need to fabricate two pinning centers,
between which the wall hops each time the AC pulse is applied. 


GT thanks Monka-shou,
 Japan and The Mitsubishi Foundation for financial support.


\end{document}